\documentstyle[11pt,titlepage,psfig]{article}

\begin{document}
{\Large
\centerline{ON THE CASIMIR ENERGY FOR A         }
\centerline{2$N$-PIECE RELATIVISTIC STRING        } }

\vskip 2.0cm
{\normalsize 
\centerline{    I. BREVIK$^*$                  }
\centerline{\it Division of Applied Mechanics  }
\centerline{\it Norwegian University of Science and Technology}
\centerline{\it N-7034 Trondheim, Norway   }
\centerline{                                   }
\centerline{    and                            }
\centerline{                                   }
\centerline{    R. SOLLIE$^\dagger$            }
\centerline{\it IKU, Sintef Group              }
\centerline{\it N-7034 Trondheim, Norway       } 
\centerline{                                   }
\centerline{                                   }
\centerline{                                   }
\centerline{                                   }
\centerline{   (December 1996)                      }

\vskip 0.8cm
$^*$       \hskip 1.0cm Electronic address:	Iver.H.Brevik@varme.unit.no

$^\dagger$ \hskip 1.0cm Electronic address:	Roger.Sollie@iku.sintef.no

\begin{abstract}

The Casimir energy for the transverse oscillations of a piecewise
uniform closed string is calculated. The string consists of 2$N$
pieces of equal length, of alternating type I and type II material,
and is taken to be relativistic in the sense that the velocity of
sound always equals the velocity of light. By means of a new recursion
formula we manage to calculate the Casimir energy for arbitrary
integers $N$. Agreement with results obtained in earlier works on the
string is found in all special cases. As basic regularization method
we use the contour integration method. As a check, agreement is found
with results obtained from the $\zeta$ function method (the Hurwitz
function) in the case of low $N$ ($N$ = 1-4). The Casimir energy is
generally negative, and the more so the larger is the value of $N$. We
illustrate the results graphically in some cases. The generalization
to finite temperature theory is also given.


\end{abstract}

\section*{I. INTRODUCTION}

	When dealing with zero point energies in quantum field theory,
it is generally desirable to deal with simple models that can be
calculated explicitly, in detail. In the first place one can
demonstrate the physical equivalence between different regularization
schemes in this way.  Secondly, as a more important point physically,
such considerations can help us to understand the issue of the energy
of the vacuum state in a real system, quite a compelling goal. The
relativistic, piecewise uniform string model is in our opinion a model
that is useful in this context. It is two-dimensional, it is easy to
handle mathematically as a mechanically vibrating system, it lies open
to several different regularization schemes, and finally it is easily
generalizable to the case of finite temperatures. The model was
introduced by Brevik and Nielsen in 1990 [1], for the most simple case
of a two-piece string. Later, the model in generalized form was
analysed from various points of view [2 - 5]. We shall not give a
survey of these earlier developments here, but focus attention instead
directly on the case where the string is divided into $2N$ pieces, of
alternating type I and type II material, with $N$ an integer.  This is
the case studied in [4]. The new element in our present analysis is
that we shall show explicitly how the Casimir energy is found when $N$
is {\it arbitrary} (in [4] we carried out the calculation in full only
when $N$ = 2). The key point is that we will be able to relate a
$2(N+1)$ - piece string to a $2N$-piece string by means of a recursion
formula. The whole formalism becomes in turn remarkably simple. Again,
we see here an example of how well the composite string model fits
into the standard formalism of quantum field theory. The
regularization method that we find to be the most advantageous one, is
that involving contour integration (the so-called argument
principle). This method was introduced in the context of Casimir
calculations by von Kampen, Nijboer, and Schram [6], and was used also
in [3,4,5].  One of the virtues of this method is that it is easily
generalizable to the case of finite temperatures. We also show below
how the Casimir energy can be found, in principle, if one uses instead
the $\zeta$-function regularization (the Hurwitz function). The
equivalence is verified explicitly for the cases when $N=3$ and $N=4$
(for $N=1$ and $N=2$ the equivalence was found earlier, respectively
in [2] and in [4]).

Figure 1 shows, as an illustration, the string when $N=6$. The total
length is $L$. There are thus in this particular case 12 equal pieces,
each of length $L$/12, of alternating type I and type II material,
corresponding to tensions $T_{I}$ and $T_{II}$. The mechanical system
is for arbitrary $N$ relativistic, in the sense that the velocity of
sound everywhere equals the velocity of light:
\begin{equation}
  v_s \;=\; (T_I/\rho_I)^{1/2} 
      \;=\; (T_{II}/\rho_{II})^{1/2} \;=\; c \;,
\end{equation}
$\rho_I$ and $\rho_{II}$ being the mass densities. We shall consider
the transverse oscillations, called $\psi$, of the string. The
boundary conditions at the functions are that $\psi$ itself, as well
as the transverse elastic force $T\partial \psi/\partial \sigma$
($\sigma$ denoting the length coordinate along the string), are
continuous.

Is the any direct physical meaning of a string of this type? We are
not aware of any direct application of the model, although it seems
natural to suggest that such strings played a physical role in the
early universe. It is quite remarkable, as a result of our analysis,
that the Casimir energy is generally negative. Its absolute value
increases monotonically with $N$. That is, if there were some sort of
``phase transition'' in the early universe, a string would be able to
diminish its zero point energy by dividing itself into a larger number
of pieces. This effect is particularly transparent from the formula
for Casimir energy in the limit of $x = T_I/T_{II} \rightarrow 0;$ cf.
Eq. (38) below.

From a wider perspective, our composite string model is related to
other string models proposed in the recent past, all of them with the
main purpose of getting more physical insight into the energy spectrum
and the vacuum state. For instance, Ferrer and de la Incera analysed
the energy spectrum for an open and homogeneous string, with charges
attached to its ends, in a magnetic background [7]. See also related
papers of Odintsov, Lichtzier, and Bytsenko [8], and of Odintsov
[9]. The connection between the Casimir energy phenomena and tachyon
problems were studied by Nesterenko [10], and by D'Hoker, Sikivie, and
Kanev [11]. Anm interesting variant of the composite string model is
to assume a {\it twisted} string loop; cf. the recent paper of Bayin,
Krisch, and Ozcan [12].

We put henceforth $\hbar= c = 1$. The next section deals with the
contour integration method, for the case of an arbitrary integer
$N$. The central recursion formula is given in Eq. (11); its solution
is given in Eq. (15). Sec. III deals with diagonalization of the
elemental matrix $\bf \Lambda$, and derives essentially the dispersion
function. The basic integral expression for the Casimir energy is
derived in Sec. IV.  Sec. V deals with the alternative $\zeta$
function regularization technique. The generalization to finite
temperature theory is given in Sec. VI.  Conclusions are given in
Sec. VII.

\section*{II. GENERAL FORMALISM}

\subsection*{A. The eigenvalue problem}

 The string of total length $L$ is assumed to be divided into $2N$
pieces, of alternating type I and type II material as mentioned above,
and has thus $2N$ junctions which will be numbered by $j$ = 1, 2,...,
$2N$. We introduce the symbol $x$ for the tension ratio, and also the
symbol $p_N$ (cf.  [4]):
\begin{equation}
  x   = T_I/T_{II}    \;,  \kern 1.0cm
  p_N = \omega L /N   \;.
\end{equation}
It is convenient to introduce also another symbol $\alpha$:
\begin{equation}
 \alpha \;=\; (1-x)/(1+x) \;.
\end{equation}
As shown in Ref. [4], the eigenfrequencies $\omega$ of the string are
determined from the equation
\begin{eqnarray}
  \mbox{det}\Bigl[{\bf M}_{2N}(x,p_N) - {\bf 1} \Bigr] = 0 \;,
\end{eqnarray}
where
\begin{eqnarray}
  {\bf M}_{2N}(x,p_N)  &\;=\;&    \prod_{j=1}^{2N} {\bf M}^{(j)}(x,p_N)  \;.
\end{eqnarray}

The component matrices can be expressed as
\begin{eqnarray}
 {\bf M}^{(j)}(x,p_N)  \;=\; 
 \cases{ {1+x \over 2x}    
    \left( \begin{array}{cc}      1            &  -\alpha\,e^{-ijp_N} \\
                           -\alpha\,e^{ijp_N}  &           1
           \end{array}  \right)  \;,  &  $j$ odd  \cr
& \cr
& \cr
         {1+x \over 2 }    
    \left( \begin{array}{cc}     1             &   \alpha\,e^{-ijp_N} \\
                            \alpha\,e^{ijp_N}  &           1
           \end{array}  \right)  \;,  &  $j$ even }
\end{eqnarray}
for $j$ = 1, 2,..., $(2N-1)$. At the last junction, for $j$ = $2N$,
the matrix will be of a particular form (here and henceforth given an
extra prime for clarity):
\begin{eqnarray}
  {\bf M}'^{(2N)}(x,p_N)  &\;=\;&  {1+x \over 2}    
    \left( \begin{array}{cc}  e^{-iNp_N}       &   \alpha\,e^{-iNp_N} \\
                            \alpha\,e^{iNp_N}  &     e^{iNp_N}
           \end{array}  \right) \;.
\label{Mprime}
\end{eqnarray} 
From (5) - (7) it follows that, except from a scaling factor, the
matrix ${\bf M}_{2N}$ will depend on $x$ only through the variable
$\alpha(x)$.  It is therefore convenient to scale the matrices as
${\bf M}_{2N}(x,p_N) = [(1+x)^2/4x]^N {\bf m}_{2N}(\alpha,p_N)$.  The
new matrices can be calculated as
\begin{eqnarray}
 {\bf m}_{2N}(\alpha,p_N)  
      &\;=\;&    \prod_{j=1}^{2N} {\bf m}^{(j)}(\alpha,p_N)  \;,
\label{mproduct}
\end{eqnarray}
where
\begin{eqnarray}
 {\bf m}^{(j)}(\alpha,p_N)  \;=\; 
  \left( \begin{array}{cc}      1              &  \mp \alpha\,e^{-ijp_N} \\
                        \mp \alpha\,e^{ijp_N}  &            1
            \end{array}  \right)     \;
\end{eqnarray}
for $j$ = 1, 2,..., $(2N-1)$. Here the sign convention is to use +/--
for even/odd $j$. The last matrix ${\bf m}'^{(2N)}(\alpha,p_N)$ in (8)
can be read off directly from (7).

\subsection*{B. Exact solution for arbitrary $N$}

	We shall now calculate the matrix ${\bf m}_{2N}(\alpha,p_N)$
for general $N$. This aim will be achieved by first establishing a
recursion formula. For a string that is divided into $2(N + 1)$ pieces
we can, according to (8), write the scaled resultant matrix as
\begin{eqnarray}
 {\bf m}_{2(N+1)}  &=&
\Bigl[ {\bf m}^{(1)}  \cdot \dots \cdot  {\bf m}'^{(2N)} \Bigr] \cdot  
\Bigl[({\bf m}'^{(2N)})^{-1} \cdot
   {\bf m}^{(2N)}  \cdot    {\bf m}^{(2N+1)}      \cdot
   {\bf m}'^{(2N+2)}  \Bigr]          \;. \nonumber \\
\end{eqnarray}
All these matrices have $p_{N+1}$ as their second argument. We can
therefore write
\begin{eqnarray}
   {\bf m}_{2(N+1)}(\alpha,p_{N+1})  &=&
   {\bf m}_{2N    }(\alpha,p_{N+1}) \cdot {\bf \Lambda}(\alpha,p_{N+1})  \;,
\end{eqnarray}
where the matrix ${\bf \Lambda}$ is a product of four matrices, ${\bf
\Lambda} = ({\bf m}'^{(2N)})^{-1} \cdot {\bf m}^{(2N)} \cdot {\bf
m}^{(2N+1)} \cdot {\bf m}'^{(2N+2)}$ , evaluated at $p_{N+1}$. We find
that
\begin{eqnarray}
 {\bf \Lambda}(\alpha,p)  \;=\; 
     \left( \begin{array}{cc}    a     &  b       \\
                                 b^*   &  a^*     
            \end{array}  \right)                  \;,
\end{eqnarray}
where
\begin{eqnarray}
  a &=&  e^{-ip} -  \alpha^2          \;,  \\
  b &=&  \alpha(e^{-ip} - 1)          \;.
\end{eqnarray}
It is seen that the matrix ${\bf \Lambda}$ does not depend on $N$
explicitly, but only through the variable $p = p_{N+1} = \omega L
/(N+1)$.  This fact will enable us to give an explicit solution, since
then
\begin{eqnarray}
 {\bf m}_{2N}(\alpha,p_N)  &=&  {\bf \Lambda}^N(\alpha,p_N)  \;.
\end{eqnarray}
The obvious way to continue is now to calculate the eigenvalues of
${\bf \Lambda}$, and express the elements of ${\bf M}_{2N}$ as powers
of these. Before doing that, we will check the formalism for low
values of $N$.

\subsection*{C. The case $N$ = 1}

	This is the trivial case, since
\begin{eqnarray}
 {\bf M}_{2}  &=&      
     \left( \begin{array}{cc}    M_{11}   &  M_{12}   \\
                                 M_{21}   &  M_{22}     
            \end{array}  \right)      
  \;=\;       
      {(1+x)^2 \over 4x}
     \left( \begin{array}{cc}    a     &  b       \\
                                 b^*   &  a^*     
            \end{array}  \right)     \;, 
\end{eqnarray}
which means that
\begin{eqnarray}
  M_{11} &=&  {(1+x)^2 \over 4x } \Biggl[
              e^{-i\omega L} - \Bigl({1-x \over 1+x} \Bigr)^2 \Biggr] \;,  \cr
  M_{12} &=&  {1-x^2 \over 4x }   \Bigl(e^{-i\omega L} - 1 \Bigr)     \;,
\end{eqnarray}
since $p$ is here equal to $\omega L$.

\subsection*{D. The case $N$ = 2}

Using the general formalism, we find that
\begin{eqnarray}
 {\bf M}_{4}  &=&      
     \left( \begin{array}{cc}    M_{11}   &  M_{12}    \\
                                 M_{21}   &  M_{22}     
            \end{array}  \right)      
  \;=\;       
        {(1+x)^4 \over 16x^2}
     \left( \begin{array}{cc}    a^2+bb^*     &  b(a+a*)  \\
                                 b^*(a+a^*)   &  {a^*}^2+bb^*     
            \end{array}  \right)     \;, 
\end{eqnarray}
which means that
\begin{eqnarray}
  M_{11} &=&  {(1+x)^4 \over 16x^2 } 
              \Bigl[(e^{-ip}-\alpha^2)^2+2\alpha^2 (1-\cos p) \Bigr] \;,  \cr
  M_{12} &=&  {(1+x)^2(1-x^2) \over 8x^2 }  
              \Bigl(e^{-ip} - 1 \Bigr)(\cos p -\alpha^2) \;,
\end{eqnarray}
with $p=\omega L /2$. These expressions are in agreement with
Eqs. (15) in Ref. [4]. Our present notation implies a significant
simplification in the expressions, as compared to those given in [4].

\subsection*{E. The case $N$ = 3}

This situation, corresponding to a six-piece string, can be analysed
similarly. We obtain
\begin{eqnarray}
 {\bf M}_{6}  &=&  {(1+x)^6 \over 64x^3}
   \left( \begin{array}{cc}   
      a(a^2+bb^*) + bb^*(a+a^*)            &  b\bigl[a^2+bb^*+a^*(a+a^*)\bigr]  \\
      b^*\bigl[{a^*}^2+bb^*+a(a+a^*)\bigr] &  a^*({a^*}^2+bb^*) + bb^*(a+a^*)     
          \end{array}  \right)     \;.  \nonumber \\
\end{eqnarray}

\section*{III.  DIAGONALIZATION}

 We have so far found the exact solution (15) for ${\bf m}_{2N}$.  To
calculate powers of ${\bf \Lambda}$, we will diagonalize this
matrix. First, we see that the eigenvalues $\lambda_\pm$ of ${\bf
\Lambda}$ are roots of the polynomial
\begin{eqnarray}
 P(\lambda) \;=\;   \mbox{det}({\bf \Lambda}-\lambda {\bf 1}) 
            \;=\;   \lambda^2-2(\cos p -\alpha^2)\lambda+(1-\alpha^2)^2  \;,
\end{eqnarray}
giving
\begin{eqnarray}
 \lambda_\pm  \;=\;   \cos p - \alpha^2 
          \pm     \Bigl[\Bigl   (\cos p - \alpha^2)^2 
                              - ( 1     - \alpha^2)^2 \Bigr]^{1/2}  \;.
\end{eqnarray}
These eigenvalues are in general complex. Powers of the matrix ${\bf
\Lambda}$ are
\begin{eqnarray}
 {\bf \Lambda}^N  &=&  {\bf K}      
      \left( \begin{array}{cc}   
          \lambda_+^N  &  0   \\
               0       &  \lambda_-^N
            \end{array}  \right)  {\bf K}^{-1}    \;,     
\end{eqnarray}
where $\bf K$ is a matrix whose columns consist of the eigenvectors of
${\bf \Lambda}$.  Also note that we can calculate the determinant and
the trace of ${\bf \Lambda}^N$ directly; from (23) we get
\begin{eqnarray}
 \mbox{det}( {\bf \Lambda}^N )  =  \lambda_+^N   \lambda_-^N    \;, \kern 1.0 cm
 \mbox{tr} ( {\bf \Lambda}^N )  =  \lambda_+^N + \lambda_-^N    \;.
\end{eqnarray}
Let us now consider the dispersion function for the string. According
to (4), the dispersion function is essentially the same as
$\mbox{det}({\bf M}_{2N} - {\bf 1})$.  Let us denote the latter
function by $G_N^x(\omega)$:
\begin{equation}
   G_N^x(\omega) \;=\; \mbox{det}({\bf M}_{2N} - {\bf 1}) 
             \;=\; \mbox{det}({\bf M}_{2N}) - \mbox{tr}({\bf M}_{2N}) + 1 \;.
\end{equation}
We can then write ${\bf M}_{2N} = (1-\alpha^2)^{-N} {\bf \Lambda}^N$. It follows that
\begin{eqnarray}
   G_N^x(\omega) &=& (1-\alpha^2)^{-2N} \lambda_+^N \lambda_-^N
              -  (1-\alpha^2)^{-N} (\lambda_+^N + \lambda_-^N) + 1  \nonumber  \\   
             &=& 2- (1-\alpha^2)^{-N} (\lambda_+^N + \lambda_-^N)   \;,
\end{eqnarray}
in view of the relationships $\lambda_+ \lambda_-= (1-\alpha^2)^{2}$,
$\lambda_+ + \lambda_-= 2(\cos p-\alpha^2)$ which follow from (21).
Although the eigenvalues $\lambda_\pm$ are in general complex, as
mentioned, the combinations $\lambda_+ \lambda_-$ and $(\lambda_+ +
\lambda_-)$ are always real. Starting from the expression (26), we can
now calculate the Casimir energy of the system.

\section*{IV. CASIMIR ENERGY. CONTOUR INTEGRATION METHOD}

	The Casimir energy $E_N$ describes the effect from the
nonhomogeneity of the string only, and is thus required to vanish for
a uniform string. Therefore, $E_N$ is equal to the zero- point energy
for the composite string, minus the zero-point energy $E_N^{I+II}$ for
the uniform string, i.e.,
\begin{equation}
     E_N  \;=\; E_N^{I+II}  \;-\;  E_{uniform}  \;.
\label{energy}
\end{equation} 
Because the string is assumed to be relativistic, satisfying the
condition (1), it is irrelevant here whether the uniform string is
made up of type I, or type II, material. The energy $E_{uniform}$ is
the same in either case (cf. also the discussion on this point in
Ref. [1]).

The most general and powerful way to proceed in the present case is to
use the contour integration method. The starting point is the
so-called argument principle, as used also in previous works [3-5]:
\begin{equation}
    {1 \over 2\pi i}
        \oint \omega {d \over d\omega} \ln \vert g(\omega) \vert \,d\omega 
    \;=\; \sum \omega_0 - \sum \omega_\infty  \;.
\end{equation}
Here $g(\omega)$ is any meromorphic function whose zeros are
$\omega_0$ and whose poles are $\omega_\infty$ inside the integration
contour. We choose the contour shown in Fig. 2, and identify
$g(\omega)$ with the dispersion function $g_N^x(\omega)$ for the
string. The last-mentioned function is essentially the same as our
function $G_N^x(\omega)$ above, defined in Eq. (25), but we will have
to introduce a modifying $x$-dependent factor between $g_N^x(\omega)$
and $G_N^x(\omega)$ to satisfy the limiting constraint on the system
(see below). Before determining this factor, let us in accordance with
(27) subtract off the zero-point energy of the uniform string,
corresponding to $x=1$:
\begin{equation}
   E_N(x) \;=\; {1 \over 4\pi}
        \oint \omega {d \over d\omega} 
        \ln \left| { g_N^x(\omega)  \over g_N^{x=1}(\omega)}   \right| \,d\omega  \;.
\end{equation}
The contribution to the integral from the semicirle in Fig. 2 is seen
to vanish in the limit $R \rightarrow \infty$. The remaining integral
along the imaginary frequency axis $(\xi = -i \omega$) is integrated
by parts, while keeping $R$ finite and taking advantage of the
symmetry of the integrand about the origin.  We get
\begin{equation}
   E_N(x) = - {R \over 2\pi} \,
     \ln \left| { g_N^x(iR)  \over g_N^{x=1}(iR)}  \right|    
        +  {1 \over 2\pi}
     \int_0^R  \ln \left| 
      { g_N^x(i\xi)  \over g_N^{x=1}(i\xi)}  \right|  \,d\xi  \;.
\end{equation}
The constraint that we shall impose on the system is that the surface
term vanishes in the limit of large $R$, i.e.,
\begin{equation}
 \lim_{R \rightarrow \infty} 
     {  g_N^{x}(iR)  \over g_N^{x=1}(iR) } \;=\; 1.
\label{calN}
\end{equation}
From (22) it follows, with $p = iRL/N$, $RL/N$ being a large quantity,
that
\begin{eqnarray}
   \lambda_+  \simeq  e^{RL/N} -2 \alpha^2    \;,  \kern 0.8cm
   \lambda_-  =  {\cal O}(e^{-RL/N})      \;.
\end{eqnarray}
Then (26) yields
\begin{eqnarray}
    G_N^x(iR)      \simeq  -  (1-\alpha^2)^{-N} \, e^{RL}  \;,  \kern 0.8cm
    G_N^{x=1}(iR)  \simeq  -           e^{RL}  \;,
\end{eqnarray}
and it follows that the sought relationship between $G_N^x(\omega)$
and the dispersion function $g_N^x(\omega)$ is
\begin{eqnarray}
    g_N^x(\omega)  = (1-\alpha^2)^N G_N^x(\omega) 
               = 2(1-\alpha^2)^N - (\lambda_+^N + \lambda_-^N).
\end{eqnarray}
The condition (31) is thereby satisfied.

	To calculate the integral in (30) we need to know also
$g_N^{x=1}(i\xi)$. In this case $\alpha=0$, and one can easily show
that $\lambda_\pm = \exp(\pm q)$, with $q = \xi L /N$. One finds that
\begin{eqnarray}
   g_N^{x=1}(i\xi) &=&   -4 \sinh^2 ({Nq \over 2}) \;,
\end{eqnarray}
and so we arrive at the following expression for the Casimir energy,
for arbitrary $x$ and an arbitrary integer $N$,
\begin{eqnarray}
   E_N(x) &=&  {N  \over 2\pi L}
        \int_0^\infty  \ln \left| 
      { 2(1-\alpha^2)^N - [\lambda_+^N(iq) + \lambda_-^N(iq)] 
      \over   4 \sinh^2(Nq/2) }  \right| \,dq  \;.
\label{CasPrep}
\end{eqnarray}
Here the eigenvalues $\lambda_\pm$ for complex arguments are
\begin{eqnarray}
 \lambda_{\pm}(iq) &=& \cosh q - \alpha^2 
    \pm \Bigl[ (\cosh q - \alpha^2)^2 - (1-\alpha^2)^2 \Bigr]^{1/2}   \;.
\end{eqnarray}
Figure 3 shows how $E_N L$ (in dimensional units $E_N L/\hbar c$)
varies with $x$ for some different values of $N$. Since $\alpha$
occurs quadratically in (26) and (22), it follows that the eigenvalue
spectrum of the system is invariant under the transformation $x
\rightarrow 1/x$. It is therefore sufficient to show the variations in
Casimir energy for the tension ratio interval $0 < x \leq 1$ only. It
is seen that the energy is generally negative, and the more so the
larger is the integer $N$. A string, initially uniform corresponding
to Casimir energy equal to zero, can accordingly at any time diminish
its zero-point energy simply by dividing itself into a larger number
of pieces of alternating type I/II material. It becomes very natural
to wonder if not ``phase transitions'' of this sort were playing a
physical role at some stage in the early universe.

Let us consider finally the limiting case of extreme tension ratio, $x
\rightarrow 0$. It turns out that this case is solvable
analytically. Namely, since now $\alpha \rightarrow 1$, we see from
(37) that $\lambda_- = 0$, $\lambda_+ = 4 \sinh^2(q/2)$. From (36) we
then get
\begin{eqnarray}
 E_N(0) &=& {N  \over \pi L}
            \int_0^\infty  \ln \left| 
      { 2^{N} \sinh^{N}(q/2) 
      \over   2 \sinh(Nq/2) }  \right| \,dq  
       \;=\; -{\pi \over 6L} (N^2-1)    \;. 
\label{CasZero}
\end{eqnarray}
This is quite a remarkable result.  In this limiting case the Casimir
energy is, apart from an additive constant, simply quadratic in
$N$. For $N$ = 1, the energy vanishes, in accordance with Eq. (22) in
Ref. [1].  For $N$ = 2, the energy becomes $E_2(0) = -\pi/2L$, in
accordance with Eq. (27) in Ref. [4]. In Figure 4 we have plotted the
Casimir energy, normalized with the energy $E_N(0)$ at $x=0$, i.e.,
$-E_N(x)/E_N(0)$. The figure displays all the different values of $N$
shown in Figure 3. Note that, within numerical accuracy, all the
curves seem to collapse into one single curve. This suggests a very
simple scaling of the energy with $N$, for arbitrary values of $x$.

\section*{V. $\zeta$ - FUNCTION METHOD}

	This powerful regularization method (for a general treatise,
see Ref. [13]) can be used as an alternative to calculate the Casimir
energy. We then first have to calculate the eigenvalue spectrum for
$\omega$ explicitly, by solving the dispersion relation $G_N^x(\omega)
= 0$ (or $g_N^x(\omega)=0$).  From (26) it follows that we have to
solve the equation
\begin{eqnarray}
         \lambda_+^N + \lambda_-^N = 2(1-\alpha^2)^N
\end{eqnarray}
with respect to $\omega$, $\lambda_\pm$ being given by (22). It is
convenient to make use of the following simple recursion formula for
$S(N) \equiv \lambda_+^N + \lambda_-^N$:
\begin{eqnarray}
   S(N) = 2(\cos p -\alpha^2) S(N-1) - (1-\alpha^2)^2 S(N-2)\;, 
   \kern 1.0cm N \geq 2.
\end{eqnarray}
This formula follows directly from (22). It is assumed here that the
value of $p$ is kept fixed, equal to $\omega L/N$, at all the
recursion steps of (40). The initial values of $S(N)$ are $S(0) = 2$,
$S(1) = \lambda_+ + \lambda_- = 2(\cos p -\alpha^2)$.

Once the eigenfrequency spectrum has become determined for some chosen
value of $N$, the zero-point energy can be calculated as
$\hbox{$\scriptstyle{ 1\over 2 }$} \sum \omega_n$, summed over all the
branches. The degeneracy has to be taken into account explicitly, for
each branch. The $\zeta$ function that comes into play here is the
Hurwitz function, originally defined as
\begin{eqnarray}
   \zeta(s,a) \;=\; \sum_{n=0}^\infty (n+a)^{-s}
  \kern 2.0cm \mbox{( $0 < a \leq 1$,  Re $s > 1$)} \;.
\end{eqnarray}
In practice, we need only to take into account the property
\begin{eqnarray}
   \zeta(-1,a) \;=\; -\hbox{$\scriptstyle{  1\over 2 }$} 
                (a^2-a + \hbox{$\scriptstyle{  1\over 6 }$})
\end{eqnarray}
of the analytically continued function (cf. Ref. [2]).

Let us check the $\zeta$-function method in some simple cases. First,
if $N=1$ we see from (39) that $\cos p = \cos \omega L = 1$, which
means $\omega = 2 \pi n/L$ with $n$ = 1, 2, 3,... . This is the same
spectrum as for a uniform string. The Casimir energy accordingly
vanishes, as it should. Next, if $N=2$ we find from (39) and (40) that
there are two branches, given by $\cos p = 1$ and $\cos p = 2\alpha^2
- 1$ respectively, with $p = \omega L/2$. This is in agreement with
Eq. (21) in Ref. [4], and will not be further considered here. Let us
instead put $N=3$. Then (39) and (40) lead to the equation
\begin{eqnarray}
  (\cos p - 1)(2 \cos p - 3\alpha^2 + 1)^2 \;=\; 0 
\end{eqnarray}
for determining the allowed values of $p = \omega L/3$.  There are
thus two branches in this case. The first branch, corresponding to
$\cos p=1$, is degenerate. One may physically associate this
degeneracy with the right-left symmetry of the uniform string. The
second branch, corresponding to $\cos p = \hbox{$\scriptstyle{ 1\over
2 }$}(3\alpha^2 - 1)$, is also degenerate. Mathematically, this
degeneracy occurs because of the second power of the second factor in
(43). Physically, the degeneracy may be considered as a consequence of
two single branches that have merged together. The solution of the
second branch can be written
\begin{eqnarray}
 \omega  \;=\;  {3     \over L} \arccos {3\alpha^2 -1 \over 2}
     \;=\;  {3 \pi \over L} \times
 \cases{ (\beta + 2n) \;,     \cr
         (2-\beta+2n) \;,   }
\end{eqnarray}
where $n$ = 0, 1, 2,..... and where $\beta$ is a number lying in the
interval $0 < \beta \leq 2/3$. The zero-point energy of the
composite string becomes
\begin{eqnarray}
E_3^{I+II} &=&  2 \times \hbox{$\scriptstyle{  1\over 2 }$}
            \Bigl({6\pi \over L} \Bigr) \sum_{n=1}^\infty n  \nonumber \\
           &+&  2 \times \hbox{$\scriptstyle{  1\over 2 }$}
            \Bigl({3\pi \over L} \Bigr) \sum_{n=0}^\infty (2n+\beta)
            +   2 \times \hbox{$\scriptstyle{  1\over 2 }$}
            \Bigl({3\pi \over L} \Bigr) \sum_{n=0}^\infty (2n+2-\beta)
  \;,
\end{eqnarray}
the prefactor 2 in each term describing the degeneracy. It is seen
that it is the presence of the $\beta$-term that forces us to use the
Hurwitz function instead of the Riemann function. Use of the
relationship (42) now leads to
\begin{eqnarray}
E_3^{I+II} \;=\; { 6\pi \over L} \zeta(-1,1)
             +   {12\pi \over L} \zeta(-1,\beta/2)
           \;=\; - {3\pi \over 2L} (1-\beta)^2 \;,
\end{eqnarray}
where we have taken into account that $\zeta(-1,1-a)=\zeta(-1,a)$.
Subtraction of $E_{uniform}=-\pi/6L$ yields the Casimir energy for
$N=3$:
\begin{eqnarray}
  E_3(x) \;=\; { \pi \over 6L} \Bigl[ 1-9(1-\beta)^2 \Bigr]  \;.
\end{eqnarray}
Explicit calculation shows that this expression gives values in
agreement with those found from (36). In particular, $E_3(0) =
-4\pi/3L$, in agreement with (38).

Finally, let us put $N=4$. From (39) and (40) we then obtain
\begin{eqnarray}
  (\cos p - 1) (\cos p - \alpha^2)^2 (\cos p - 2\alpha^2 + 1) \;=\; 0 
\end{eqnarray}
as the equation determining $p=\omega L/4$. There are in this case
three branches. The two first branches, corresponding to $\cos p = 1$
and $\cos p = \alpha^2$, are degenerate, whereas the third branch,
corresponding to $\cos p = 2\alpha^2-1$, is not. Let us denote the two
last-mentioned branches by indices 1 and 2. Thus, for the second
branch we have
\begin{eqnarray}
 \omega  \;=\;  {4     \over L} \arccos \alpha^2 
     \;=\;  {4 \pi \over L} \times
 \cases{ \beta_1 + 2n \;,     \cr
         (2-\beta_1+2n) \;,   }
\end{eqnarray}
with $0 < \beta_1 < \hbox{$\scriptstyle{  1\over 2 }$}$, whereas for the third branch
\begin{eqnarray}
 \omega  \;=\;  {4     \over L} \arccos (2\alpha^2 - 1) 
     \;=\;  {4 \pi \over L} \times
 \cases{ \beta_2 + 2n \;,     \cr
         (2-\beta_2+2n) \;,   }
\end{eqnarray}
with $0 < \beta_2 \leq 1$. The zero-point energy becomes
\begin{eqnarray}
E_4^{I+II}  &=&  2 \times \hbox{$\scriptstyle{  1\over 2 }$}
             \Bigl({8\pi \over L} \Bigr) \sum_{n=1}^\infty n \nonumber \\
            &+&  2 \times \hbox{$\scriptstyle{  1\over 2 }$}
             \Bigl({4\pi \over L} \Bigr) \sum_{n=0}^\infty (2n+\beta_1)
             +   2 \times \hbox{$\scriptstyle{  1\over 2 }$}
             \Bigl({4\pi \over L} \Bigr) \sum_{n=0}^\infty (2n+2-\beta_1)
 \cr
            &+&           \hbox{$\scriptstyle{  1\over 2 }$}
             \Bigl({4\pi \over L} \Bigr) \sum_{n=0}^\infty (2n+\beta_2)
             +            \hbox{$\scriptstyle{  1\over 2 }$}
             \Bigl({4\pi \over L} \Bigr) \sum_{n=0}^\infty (2n+2-\beta_2)
 \cr
            &=&  { 8\pi \over L} \zeta(-1,1)
             +   {16\pi \over L} \zeta(-1,\beta_1/2)
             +   { 8\pi \over L} \zeta(-1,\beta_2/2)   \cr
            &=& -{\pi \over L} [2(1-\beta_1)^2 + (1-\beta_2)^2
                - \hbox{$\scriptstyle{  1\over 3 }$}] \;,
\end{eqnarray}
and so the Casimir energy becomes
\begin{eqnarray}
  E_4(x) \;=\; { \pi \over 2L} 
           \Bigl[ 1-4(1-\beta_1)^2 -2(1-\beta_2)^2\Bigr]  \;.
\end{eqnarray}
Again, explicit evaluation leads to agreement with the integral
formula (36). We see that when using the $\zeta$-function method we
have to determine the eigenfrequency spectrum explicitly, and
thereafter put in the degeneracies by hand. The very useful bonus
associated with our contour integration technique above, is that the
degeneracies precisely correspond to the multiplicities of the zeros
in the argument principle, Eq. (28), and need not be taken into
account explicitly. In view of these properties, the contour
integration method appears to be the simplest method in the present
case.

\section*{VI. FINITE TEMPERATURE THEORY}

	The generalization of the T = 0 theory above to the case of
finite temperatures is readily accomplished by starting from the
integral expression (36) and replacing the integral over imaginary
frequencies by a sum:
\begin{eqnarray}
  \int_0^\infty d\xi \; \rightarrow \; 
        2\pi k_B T \sum_{n=0}^{\infty} {'} \kern 0.8cm \;,
\end{eqnarray}
the prime meaning that the $n=0$ term is taken with half
weight. Introducing the Matsubara frequencies $\xi_n = 2\pi n k_B T$
we then get
\begin{eqnarray}
   E_N^T(x) &=&  k_B T \sum_{n=0}^{\infty} {'}
      \ln \left| 
      { 2(1-\alpha^2)^N - [\lambda_+^N(i\xi_n L/N) + \lambda_-^N(i\xi_n L/N)] 
      \over   4 \sinh^2(\xi_n L/2) }  \right|    \nonumber \\ 
\end{eqnarray}
as the expression giving the Casimir energy, valid at any temperature
$T$. Here, $\lambda_\pm(i\xi_n L/N)$ are given by (37), with $q
\rightarrow q_n= \xi_n L/N$. It is useful to note that
\begin{eqnarray}
  \lambda_+(iq_n) + \lambda_-(iq_n) = 2(\cosh q_n -\alpha^2) \;.
\end{eqnarray}

There are several special cases of interest here. First, if the string
is uniform ($x$ = 1), we get from (54) that $E_N^T(1)=0$. This is as
we would expect, since even at finite temperatures the Casimir energy
is intended to describe the influence from the inhomogeneity of the
string only. Next, if $N$ = 1, $x$ arbitrary, we also get a vanishing
result, $E_1^T(x)=0$. If $N$ = 2, we get for $E_2^T(x)$ the same
integral expression as in Eq. (36) in Ref. [4]. For larger values of
$N$, $N$ = 3, 4, .... , we can develop the integral expressions in the
same way. In particular, in the case of $x \rightarrow 0$ we get the
simple formula
\begin{eqnarray}
 E_N^T(0) &=& 2 k_B T  \sum_{n=0}^{\infty} {'}
     \ln \left|  { 2^{N} \sinh^{N}(\xi_n L/2N) 
      \over   2 \sinh(\xi_n L/2) }  \right| \;.
\end{eqnarray}
We shall not discuss this topic in further detail here. As shown in
Ref. [3], for practical purposes the series can be evaluated fairly
easily by means of a computer program.

	The Casimir energy found here may be helpful also in the
construction of string theory at finite temperatures (for a review,
see Chapter 8 of [13]), with further possible applications in string
cosmology.

\section*{VII. CONCLUSIONS}

	In this paper we have found the general expression for the
Casimir energy for the 2$N$-piece relativistic string, for an
arbitrary integer $N$, and for arbitrary ratio between the two kinds
of material. The expression for $E_N(x)$, at temperature $T=0$, is
given in (36).  The present work generalizes earlier works [1-4], and
is in agreement with them in all special cases. At finite temperatures
$T$, the corresponding Casimir energy $E_N^T(x)$ is given in (54).
The key new element in our analysis is the recursion formula (11) and
its explicit solution (15), which enables us to find $E_N(x)$ for
arbitrary $N$. The use of this recursion formula greatly simplifies
the calculation of the dispersion function $G_N^x(\omega)$ or
$g_N^x(\omega)$ ; cf. Eqs. (26) and (34).

The regularization method used in the derivation of (36) was the
contour integration method (argument principle), originally introduced
in Casimir-type of calculations in Ref. [6].  There are three reasons
why we consider this regularization method to be preferable in the
present problem:

\begin{itemize}
\item[1)] The method is simple, in that we do not have to solve for
the eigenvalue spectrum explicitly. Moreover, we do not have to take
into account the degeneracies explicitly; they are automatically
accounted for, in the multiplicities of the zeros in the argument
principle.

\item[2)] The generalization to arbitrary integers $N$
is straightforward.  

\item[3)] The generalization to finite temperature
theory is straightforward.  
\end{itemize}

As a check of the results obtained from the contour integration
method, we gave in Sec.V an independent derivation based upon the
$\zeta$ function method. The actual $\zeta$ function here is the
Hurwitz function. When proceeding in this way, the eigenvalue spectrum
has to be worked out, and the degeneracies have to be put in by
hand. Explicit evaluation in the cases $N=3$ and $N=4$ gave results in
agreement with the integral formula (36).

A remarkable physical result is that the Casimir energy is always
negative, and the more so the higher is the value of $N$. An eclatant
example of this behaviour is the expression (38), showing the Casimir
energy in the case $x \rightarrow 0$.  A string can always lower its
zero-point energy by dividing itself into a larger number of pieces,
of alternating type I / II material.  Perhaps were processes of this
type taking place in the early universe, as some sort of ``phase
transitions''.

	Another point worth noticing is the apparent scaling of the
Casimir energy with $N$ for arbitrary values of $x$, which is strongly
suggested by Fig. 4.

\eject
\section*{REFERENCES}

\begin{itemize}

\item[1] I. Brevik and H.B. Nielsen, Phys. Rev. D {\bf 4}1, 1185 (1990).

\item[2] X. Li, X. Shi, and J. Zhang, Phys. Rev. D {\bf 44}, 560 (1991).

\item[3] I. Brevik and E. Elizalde, Phys. Rev. D {\bf 49}, 5319 (1994). 

\item[4] I. Brevik and H.B. Nielsen, Phys. Rev. D {\bf 51}, 1869 (1995). 

\item[5] I. Brevik, H.B. Nielsen, and S.D. Odintsov, 
Phys. Rev. D {\bf 53}, 3224 (1996).

\item[6] N.G. van Kampen, B.R.A. Nijboer, and K. Schram, 
Phys. Lett. {\bf 26 A}, 307 (1968).

\item[7] E.J. Ferrer and V. de la Incera, Phys. Rev. D {\bf 52}, 1011 (1995).

\item[8] S.D. Odintsov, I.M. Lichtzier, and A.A. Bytsenko, Acta
Physica Polon. {\bf 22}, 761 (1991).

\item[9] S.D. Odintsov, Rivista del Nuovo Cimento {\bf 15}, 1 (1992).

\item[10] V.V. Nesterenko, Z. Phys. C {\bf 51}, 643 (1991).

\item[11] E. D'Hoker and P. Sikivie, Phys. Rev. Lett.{\bf 71}, 1136 (1993); 
E. D'Hoker, P. Sikivie, and Y. Kanev, Phys. Lett. B {\bf 347}, 56 (1995).

\item[12] S. S. Bayin, J. P. Krisch, and M. Ozcan, Journal of
Mathematical Physics {\bf 37}, 3662 (1996).

\item[13]  E. Elizalde, S.D. Odintsov, A. Romeo, A.A. Bytsenko and 
S. Zerbini, {\it Zeta Regularization Techniques with Applications} (World
Scientific, Singapore, 1994).

\end{itemize}

\eject
\section*{FIGURE CAPTIONS}

\begin{itemize}

\item[FIG.1] String of length $L$, of alternating type I and type II 
material, in the case when $N$ = 6.

\item[FIG.2] Integration contour in the complex $\omega$ plane.

\item[FIG.3] Nondimensional Casimir energy versus $x=T_I/T_{II}$ for 
some values of $N$.

\item[FIG.4] The Casimir energy scaled with the value at $x=0$,
plotted versus $x=T_I/T_{II}$ for the {\it same} values of $N$ as in
Fig. 3.

\end{itemize}

\eject
{\Large\bf FIGURES}
\begin{figure}[htbp]
\psfig{figure=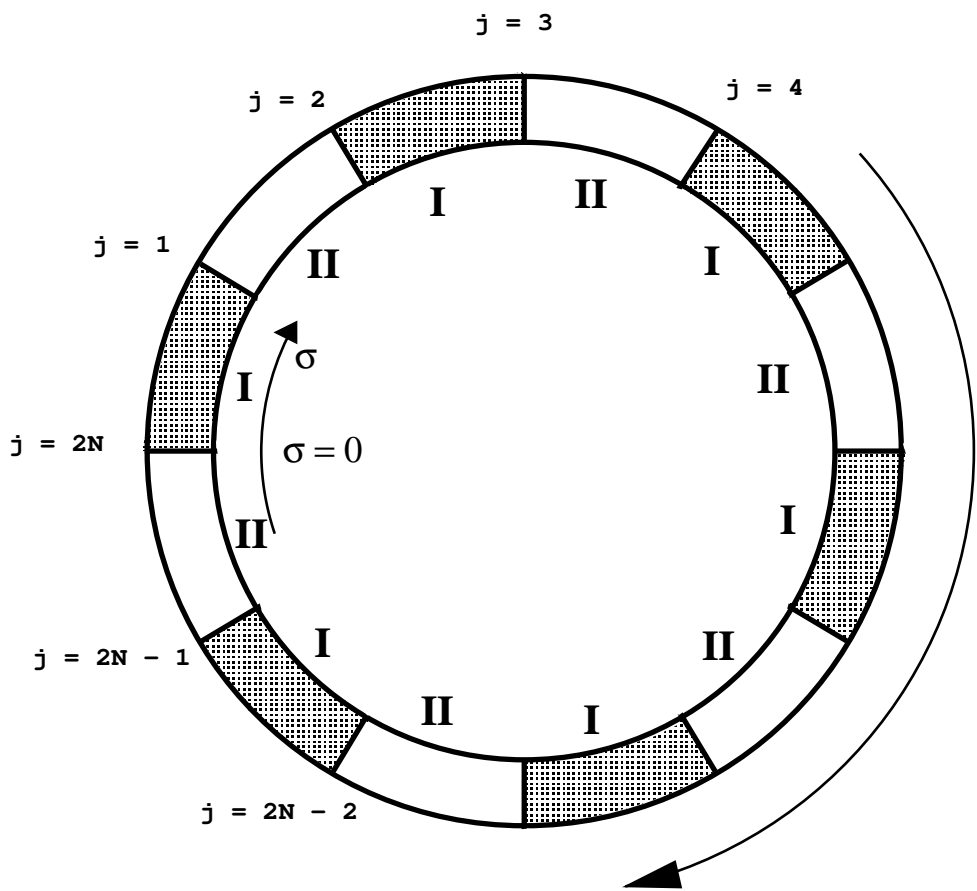,width=12cm,angle=0}
  \vskip 1.6cm
  \caption{ String of length $L$, of alternating type I and type II 
material, in the case when $N$ = 6.}
\end{figure} 

\eject
\begin{figure}[htbp]
\psfig{figure=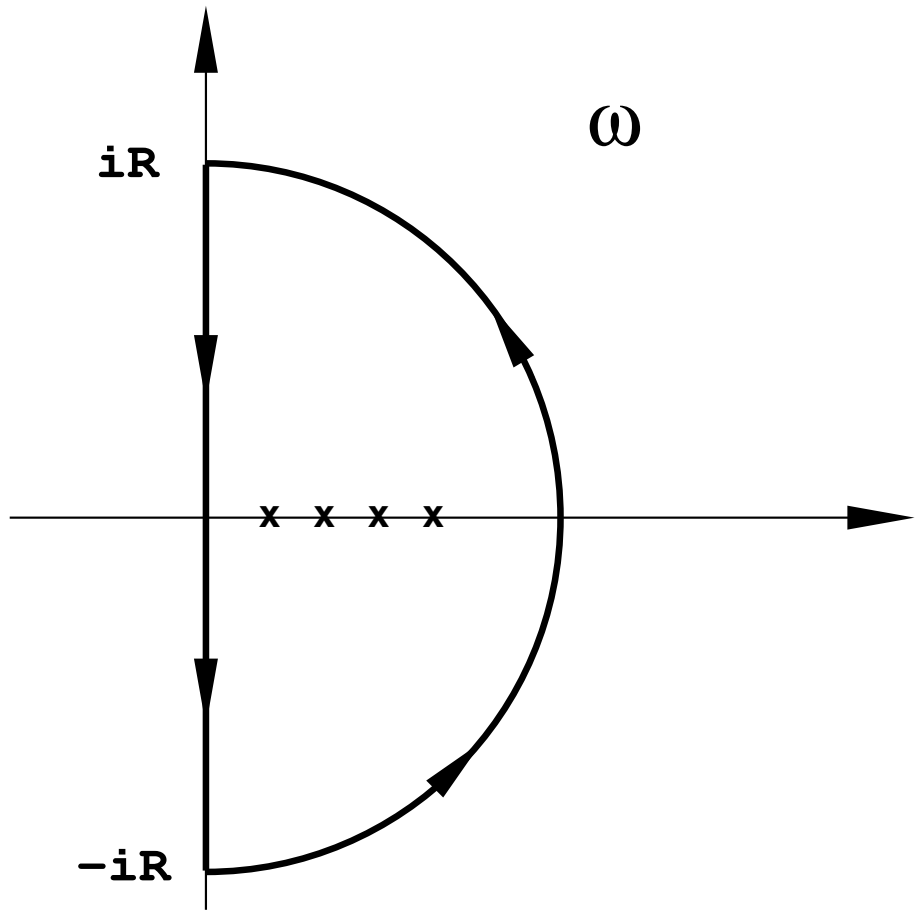,width=12cm,angle=0}
  \caption{Integration contour in the complex $\omega$ plane.}
\end{figure}

\eject
\begin{figure}[htbp]
\psfig{figure=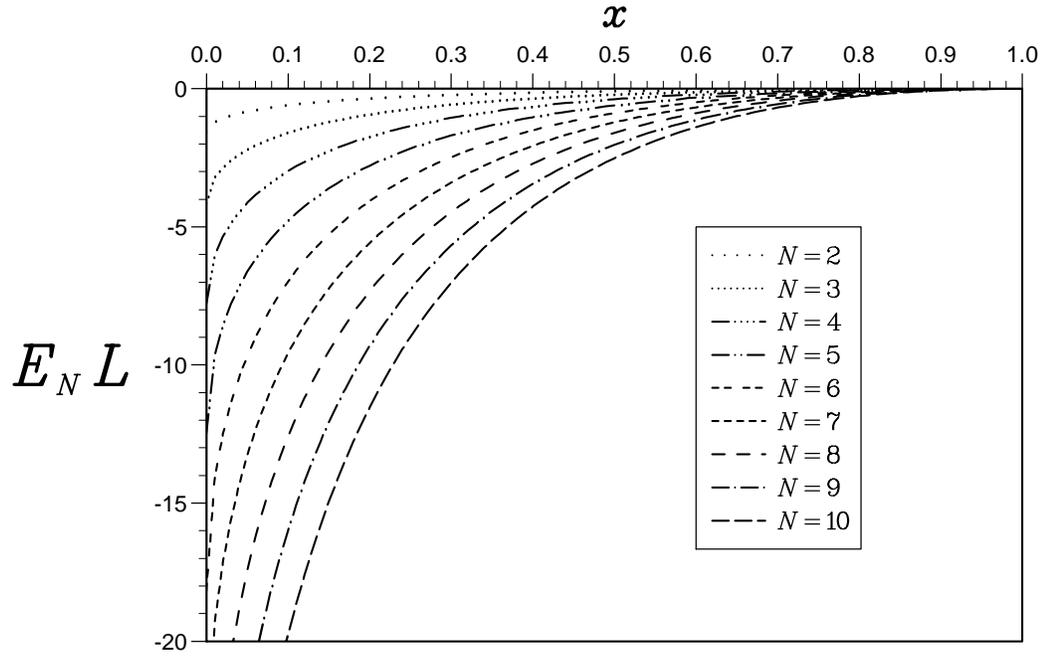,width=18cm,angle=270}
  \caption{Nondimensional Casimir energy versus $x=T_I/T_{II}$ for 
some values of $N$.}
\end{figure}

\eject
\begin{figure}[htbp]
\psfig{figure=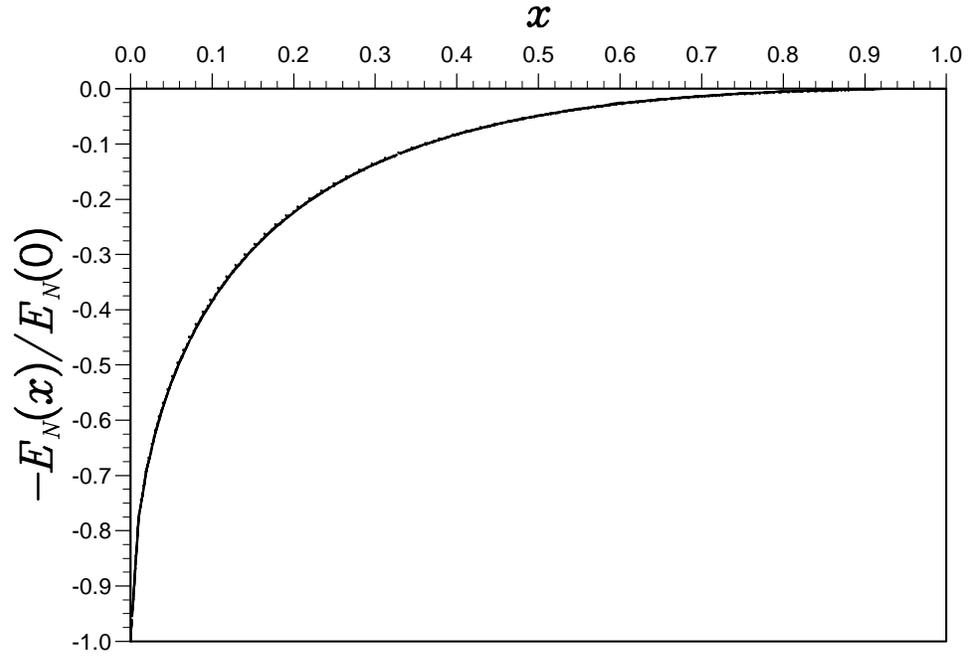,width=18cm,angle=270}
  \caption{ The Casimir energy scaled with the value at $x=0$,
plotted versus $x=T_I/T_{II}$ for the {\it same} values of $N$ as in
Fig. 3.}
\end{figure}

\end{document}